\documentstyle [preprint,aps] {revtex}
\begin{document}
\draft
\title{
	Exact Calculation of the Vortex-Antivortex
	Interaction Energy in the Anisotropic 3D XY-model
}
\author{Mahn-Soo Choi and Sung-Ik Lee}
\address{
	Department of Physics,
	Pohang University of Science and Technology,
	Pohang, 790-784, Korea
}
\maketitle

\vspace{1cm}
\begin{abstract}
We have developed an exact method to calculate the vortex-antivortex
interaction energy in
the anisotropic 3D-XY model. For this calculation, dual transformation which is
already known for the
2D XY-model was extended. We found an explicit form of this interaction energy
as a function of the anisotropic
ratio and the separation $r$ between the vortex and antivortex located on
the same layer. The form of interaction energy is $\ln r$  at the small $r$
limit
but is proportional to $r$ at the opposite limit. This form of interaction
energy
is consistent with the upper bound calculation using the variational method
by Cataudella and Minnhagen.
\end{abstract}
\pacs{PACS: 74.25.Ha,74.60.-w,75.30.Gw,75.30.Mb}

\vspace{1cm}
In studying the phase fluctuation effects of high temperature layered
superconductors,
several
approximations\cite{XY:Cataudella90,XY:Glazman90,XY:Minnhagen91a,XY:Minnhagen91b,XY:Weber91}
of the vortex-antivortex interaction energy have been suggested
at the level of the highly anisotropic 3D XY-model
(from now on called the layered XY-model) or
the Lawrence-Doniach model\cite{XY:LawDon71,XY:Bulaevskii73,XY:Clem89}.
For the bare interaction energy, Cataudella and Minnhagen\cite{XY:Cataudella90}
adopted the variational
method and found the upper bound of this energy. According to their
calculation, this
interaction energy increases linearly with the separation between the vortex
and the antivortex,
which is different from the logarithmically increasing energy of the 2D
XY-model.
This is not surprising since this energy at larger separations
should be dominated by the Josephson vortex lines connecting the
vortex and antivortex (see Fig.\ref{fig:vortexloop}).

In passing, we must clarify what we mean by a vortex-antivortex pair
on the same layer (Fig.\ref{fig:vortexloop}).
For the layered XY-model or Lawrence-Doniach model, the vortex line
cannot be disconnected, and should either be infinitely long or a closed loop.
Thus a vortex and antivortex pair on a layer should be connected by
Josephson vortex lines (Josephson strings) residing between the layers.
Recent experimental results on high temperature superconductors
\cite{XY:expa,XY:expb,XY:expc,XY:expd,XY:expe,XY:expf,XY:expg}
were, however, explained through the interpretation
that the phase fluctuations are associated with the layers and
that major roles in phase fluctuation effects are played only by the vortex and
antivortex
on the layer.
Hence the vortex-antivortex interaction energy is reasonably
defined\cite{XY:Minnhagen91a,XY:Minnhagen91b} as
the smallest energy of the very vortex loop configuration, which corresponds to
the shortest Josephson strings.
The Josephson strings just modify the interaction energy of the pair at
large separations.
Monte Carlo simulations\cite{XY:Minnhagen91a,XY:Minnhagen91b,XY:Weber91}
performed to understand the phenomenon of vortex fluctuation at finite
temperature
also support this.

In the present paper we develop a dual transform of the highly anisotropic 3D
XY-model
and obtain an  explicit form of the vortex-antivortex interaction energy as a
function of the anisotropic ratio and $r$. The dual transformation of 2D
XY-model which was developed by
Jose {\it et al.}\cite{XY:Jose77} was extended for anisotropic 3D XY-model.
This result is compared with that of the variational calculation by Cataudella
and Minnhagen\cite{XY:Cataudella90}.
We also discuss the result in connection with the high-$T_c$ superconductors.
This is under the assumption that the Lawrence-Doniach model is reduced
to the layered XY-model. The reduction is possible if the amplitude fluctuation
is strongly suppressed and the energy associated with the
induced magnetic field is neglected\cite{XY:Minnhagen91b}.

\section{Dual transformation}
We begin with the partition function
\begin{equation}
Z = \int_0^{2\pi}\{d\theta\}\;
	\exp\left[ -\sum_{ijn} S(i|j|n) \right],\
	\{d\theta\} = \prod_{ijn}d\theta_{i,j,n}
\end{equation}
\begin{equation}
-S(i|j|n) = K_\| \cos\left(\theta_{i\!+\!1,j,n}-\theta_{i,j,n}\right)
		+ K_\| \cos\left(\theta_{i,j\!+\!1,n}-\theta_{i,j,n}\right)
		+ K_\bot\cos\left(\theta_{i,j,n\!+\!1}-\theta_{i,j,n}\right)
\end{equation}
where $K_\| \equiv K$ is the intralayer coupling  constant and
$K_\bot \equiv \epsilon^2 K$ $(\epsilon \ll 1)$ is the interlayer coupling
constant.
In an approximate connection to the Lawrence-Doniach model,
\begin{equation}
K = \frac{\hbar^2}{m}\frac{|\psi|^2}{k_BT},\ \ \
\epsilon^2 = \frac{m}{M}\left(\frac{\xi_\|}{d}\right)^2,
\end{equation}
where $\psi$ is the Ginzburg-Landau order parameter,
$\xi_\|$ is the Ginzburg-Landau inplane coherence length.
The $m$ ($M$) is the effective mass parallel (perpendicular)
to the plane. After expanding each exponential factor in Fourier series
\[
e^{K\cos\phi} = \sum_{m=-\infty}^\infty I_m(K) e^{+im\phi}
	\simeq I_0(K)\sum_{m} e^{-m^2/2K} e^{+im\phi},\ (K \gg 1)
\]
where $I_m$ are modified Bessel functions,
and after integrating out $\theta_{i,j,n}$,
we can rewrite $Z$ and $S$ as
\begin{equation}
Z   \propto {\sum_{\{m\}}}'
	\exp\left[ -\sum_{ijn} S(i|j|n) \right]
\end{equation}
\begin{equation}
-S(i|j|n)       = -\frac{1}{2K}|m_x(i\!+\!1,i|j|n)|^2 -
	\frac{1}{2K}|m_y(i|j\!+\!1,j|n)|^2 -
	\frac{1}{2\epsilon^2K}|m_y(i|j|n\!+\!1,n)|^2,
\end{equation}
where the primed sum denotes the constraint
$$
m_x(i\!+\!1,i|j|n)-m_x(i,i\!-\!1|j|n) +
m_y(i|j\!+\!1,j|n)-m_y(i|j,j\!-\!1|n)
$$
\begin{equation}
\mbox{} + m_z(i|j|n\!+\!1,n)-m_z(i|j|n,n\!-\!1) = 0
\label{eqn:constraintm}
\end{equation}
for all $i$, $j$ and $n$.

The constraint for the summation over $m$ can be solved
by moving to the dual lattice of the original cubic lattice (see
Fig.\ref{fig:duallattice})
and by defining another integer field $\ell$ on the dual lattice sites
\begin{eqnarray*}
m_x(i,i\!+\!1|j|n)
    & = & \ell_z(i\!+\!\frac{1}{2}|j\!+\!\frac{1}{2}|n) -
\ell_z(i\!+\!\frac{1}{2}|j\!-\!\frac{1}{2}|n)
    -\ell_y(i\!+\!\frac{1}{2}|j|n\!+\!\frac{1}{2}) +
\ell_y(i\!+\!\frac{1}{2}|j|n\!-\!\frac{1}{2}) \\
m_y(i|j,j\!+\!1|n)
    & = & \ell_x(i|j\!+\!\frac{1}{2}|n\!+\!\frac{1}{2}) -
\ell_x(i|j\!+\!\frac{1}{2}|n\!-\!\frac{1}{2})
    -\ell_z(i\!+\!\frac{1}{2}|j\!+\!\frac{1}{2}|n) +
\ell_z(i\!-\!\frac{1}{2}|j\!+\!\frac{1}{2}|n) \\
m_z(i|j|n,n\!+\!1)
    & = & \ell_y(i\!+\!\frac{1}{2}|j|n\!+\!\frac{1}{2}) -
\ell_y(i\!-\!\frac{1}{2}|j|n\!+\!\frac{1}{2})
    -\ell_x(i|j\!+\!\frac{1}{2}|n\!+\!\frac{1}{2}) +
\ell_x(i|j\!-\!\frac{1}{2}|n\!+\!\frac{1}{2}).
\end{eqnarray*}
In terms of the field on the dual lattice, $Z$ is written in a matrix form
\begin{equation}
Z \propto \sum_{\{\ell\}} \exp\left[
		-\frac{1}{2K}\left\langle\:\ell\:\right| M
		\left|\:\ell\:\right\rangle
	\right]
\end{equation}
where the bra and ket notation is defined below
\[
\left\langle\:\ell\:\right| M \left|\:\ell\:\right\rangle
	= \sum_{\mu\nu}\sum_{ii'}\sum_{jj'}\sum_{nn'}
	\ell_\mu(i,j,n) M_{\mu\nu}(ii'|jj'|nn') \ell_\nu(i',j',n').
\]
(For explicit form of $M$, see Appendix \ref{sec:matrices}.)
Using the Poisson resummation rule
\[
\sum_{\ell=-\infty}^\infty f(\ell)
	= \sum_{q=-\infty}^\infty\int_{-\infty}^\infty d\ell\;
	 f(\ell) e^{2\pi i q\ell},
\]
it follows that
\begin{equation}
Z \propto \sum_{\{Q\}} \int\{d\ell\}\;\exp\left[
		-\frac{1}{2K}\left\langle\:\ell\:\right| M \left|\:\ell\:\right\rangle
		+ 2\pi i \left\langle\:Q(p)\:\right|\left.\:\ell\:\right\rangle
	\right].
\label{eqn:matrixformreal}
\end{equation}
where the quantity $Q_\mu (\mu = x, y\mbox{ or }z)$ is
interpreted to be the vorticity in the direction of $\mu$-axes\cite{XY:Jose77}.

Since element of the matrix $M$ depends only on the differences between the
site indices, it is convenient to calculate $Z$ in the momentum space where the
matrix
$M$ is diagonal. Then we treat only $3\times3$ matrix $M(p_x,p_y,p_z)$
\begin{equation}
Z \propto \sum_{\{Q\}} \int\{d\ell\}\;
	\exp\left[ -\int(dp) S(p) \right],\
	\int(dp) = \int_0^{2\pi}\frac{d^3p}{(2\pi)^3}
\end{equation}
with
\begin{equation}
-S(p)
	= -\frac{1}{2K} \left\langle\:\ell(p)\:\right| M(p)
	\left|\:\ell(p)\:\right\rangle
	+ 2\pi i \left\langle\:Q(p)\:\right|\left.\:\ell(p)\:\right\rangle.
\label{eqn:matrixformmomentum}
\end{equation}

Now we diagonalize the matrix $M(p)$ (Appendix \ref{sec:matrices}) with three
eigenvectors $\left|\:v_0\:\right\rangle$,
$\left|\:v_1\:\right\rangle$ and $\left|\:v_2\:\right\rangle$
and three eigenvalues $\omega_0 ( = 0)$, $\omega_1$ and $\omega_2$.
We integrate over
$\left\langle\:v_0\:\right|\left.\:\ell\:\right\rangle$,
$\left\langle\:v_1\:\right|\left.\:\ell\:\right\rangle$ and
$\left\langle\:v_2\:\right|\left.\:\ell\:\right\rangle$,
instead of $\left|\:\ell\:\right\rangle$.
Then, the zero eigenvalue gives a constraint to the vortex
configuration
$\left\langle\:Q(p)\:\right|\left.\:v_0\:\right\rangle = 0$ or
\[
Q_x(i|j\!+\!\frac{1}{2}|n\!+\!\frac{1}{2}) -
Q_x(i\!-\!1|j\!+\!\frac{1}{2}|n\!+\!\frac{1}{2}) +
Q_y(i\!+\!\frac{1}{2}|j|n\!+\!\frac{1}{2}) -
Q_x(i\!+\!\frac{1}{2}|j\!-\!1|n\!+\!\frac{1}{2})
\]
\begin{equation}
\mbox{}+ Q_z(i\!+\!\frac{1}{2}|j\!+\!\frac{1}{2}|n) -
Q_x(i\!+\!\frac{1}{2}|j\!+\!\frac{1}{2}|n\!-\!1) = 0,
\label{eqn:constraintQ}
\end{equation}
{\it i.e.}, the vortex line is either infinitely long or a closed loop.
After Gaussian integration over
$\left\langle\:v_1\:\right|\left.\:\ell\:\right\rangle$ and
$\left\langle\:v_2\:\right|\left.\:\ell\:\right\rangle$,
we finally obtain the partition function
for vortices
\begin{equation}
Z \propto {\sum_{\{Q\}}}' \exp\left[ -\int(dp)S(p) \right]
\end{equation}
with
\begin{equation}
-S(p) = -\frac{1}{2} (2\pi)^2 K \left\langle\:Q(p)\:\right|\left(
		\frac{1}{w_1}\left|\:v_1\:\right\rangle \left\langle\:v_1\:\right| +
		\frac{1}{w_2}\left|\:v_1\:\right\rangle \left\langle\:v_2\:\right|
	\right) \left|\:Q(p)\:\right\rangle
\label{eqn:finalform}
\end{equation}
where the primed sum reminds us of the constraint Eqn.(\ref{eqn:constraintQ}).

\section{Result}
The dual transformation discussed above provides an explicit form of
interaction energy of the vortex-antivortex pair on a layer
(Fig.\ref{fig:vortexloop}).
Just for convenience (noting that we are interested in the large separation
limit),
we assume that the vortex and antivortex
is placed parallel to the $y$-axes ($Q_x = 0$).
Let the separation between the vortex and the antivortex be $2 r$.
In this case, the configuration vector
$\left\langle\:Q(p)\:\right|$ is given by
\begin{equation}
\left\langle\:Q(p)\:\right| = \frac{1-e^{+i2rp_y}}{1-e^{-ip_y}}\left(
		0, -[1-e^{-ip_z}], 1-e^{-ip_y}
	\right)
\end{equation}
If we substitute this vector into (\ref{eqn:finalform}), we obtain the
interaction energy
in units of $4\pi K$ as
\begin{equation}
U(2r) = \tan^{-1}(\epsilon)\cdot(2r)
	+ \frac{2}{\pi} \int_0^{\pi/2}d\varphi
	\frac{\sin^2(2 r \varphi)}{\sin^2\varphi} F_\epsilon(\sin\varphi)
\label{eqn:interaction}
\end{equation}
where
\begin{equation}
F_\epsilon(\alpha)
	= \int_0^1 dt\left(
		\frac{1}{\sqrt{\epsilon^2 + 1-t^2}} -
		\frac{\sqrt{1-t^2}}{\sqrt{\alpha^2 + 1-t^2}\sqrt{\alpha^2+\epsilon^2 +1-t^2}}
	\right).
\label{eqn:funcF}
\end{equation}
To simplify the integral form of the interaction energy
of Eqn.(\ref{eqn:interaction}), we used the formula
\begin{equation}
\sin(2r\phi)/\sin\phi = 2\sum_{s=1}^r \cos[(2s\!-\!1)\phi].
\end{equation}
Note that in the case of $\epsilon = 0$, Eqn.(\ref{eqn:interaction}) is reduced
into
the interaction for the 2D XY-model, while even small value of $\epsilon$
induces the
term linear in $r$.

The asymptotic expansion for $r \gg 1$ (Appendix \ref{sec:asymptotic}) gives
\begin{equation}
U(2r) \simeq \tan^{-1}(\epsilon)\cdot(2r) +
	\mu_c + {\cal O}(1/r),\ \ (\mu_c = \mbox{Const.}).
\label{eqn:asymU}
\end{equation}
Cataudella and Minnhagen, in their variational calculation, took a
simple approximation, where the phases on the layers next to the
layer containing the vortex-antivortex pair are identically zero,
and found an upper bound.
Their result is expected to be larger than the values of our exact calculation.
In our notations, the interaction energy of Cataudella and Minnhagen
is as follows (in units of $4\pi K$)
\[
U_{var}(2r) \simeq \frac{\pi}{\sqrt{2}}\cdot\epsilon\cdot(2r) + \mbox{Const.}
\]
This, compared with ours
$U(2r) \simeq \left[\epsilon + {\cal O}(\epsilon)\right]\cdot(2r) +
\mbox{Const.}$,
is larger roughly by a factor $\pi/\sqrt{2}$.

For small values of $r$, simple asymptotic form is not available.
But numerical evaluation shows logarithmic increase with $2r$ of the
interaction energy (Fig.\ref{fig:intplot}(b)).
Plots of $U(2r)$ for $2r \leq 150$ and $\epsilon = 0, 0.01, 0.02, 0.03$
which include the intermediate case is shown in Fig.\ref{fig:intplot}.
It clearly shows crossover from logarithmic behavior for small $r$
to linear one for large $r$.
If the anisotropic ratio $\epsilon$ increases, the region of logarithmic
dependence decreases.
For the details, we draw $U$-$\log(2r)$ plot at Fig.\ref{fig:intplot}(b).

The modification of the bare vortex-antivortex interaction,
in consequence, leads to the correction to the two-dimensional
Berezinskii-Kosterlitz-Thouless (BKT) transition
\cite{XY:Berezi72,XY:Koster72,XY:Koster73,XY:Koster74}.
Hikami and Tsuneto\cite{XY:Hikami80} studied this effect
by the renormalization group (RG) analysis.
In their study, they assumed that the two dimensional
RG equation\cite{XY:Koster74} is valid, but subject to the cutoff
$1/\epsilon$ because the vortex-antivortex interaction energy
is logarithmic only at distances smaller than $1/\epsilon$
(in units of the lattice constant)\cite{XY:footnote}.
According to their result, for example,
the true transition temperature $T_c$ is shifted
from the two dimensional transition point $T_{BKT}$
due to the small interlayer coupling by an amount of order
$(\pi/|\ln\epsilon|)^2$.
This result has also been confirmed by the Monte Carlo simulations
\cite{XY:Minnhagen91a,XY:Minnhagen91b}.

\section{Conclusion}
We developed a dual transformation of the highly anisotropic 3D XY-model to
study
the bare vortex-antivortex interaction. We found that this dual transformation
method
provides an exact form of vortex-antivortex interaction energy as a function
of the anisotropic ratio and the separation between  the vortex-antivortex
pair.
This form of this interaction is in good agreement with a zero temperature
variational calculation by
Cataudella and Minnhagen.
The correction to the two dimensional BKT transition due to the small
interlayer coupling was also briefly discussed.

\section*{Acknowledgement}
We wish to express appreciation for the financial support of the Korean
Ministry
of Science and Technology, and the Korean Ministry of Education.

\appendix
\section{}\label{sec:matrices}
In this appendix, we summarize the informations about the matrix $M$ in
Eqn.(\ref{eqn:matrixformreal}) and Eqn.(\ref{eqn:matrixformmomentum}).
First we define some short-hand notations for convenience:
\begin{eqnarray*}
\nabla(p_\mu)    & = & 1 - \exp(-i p_\mu),\\
\Delta(p_\mu)    & = & 2(1-\cos p_\mu),\ \mu = x,y\mbox{ or } z \\
\Delta(p_x,p_y)  & = & \Delta(p_x) + \Delta(p_y) \\
\Delta(p_x,p_y,p_z)
				& = & \Delta(p_x) + \Delta(p_y) + \Delta(p_z) \\
\Delta_\epsilon(p_x,p_y,p_z)
				& = & \Delta(p_x) + \Delta(p_y) + \epsilon^2\Delta(p_z) .
\end{eqnarray*}

The Fourier transform of $M_{\mu\nu}(ii'|jj'|nn')$ is defined as
\begin{equation}
M_{\mu\nu}(ii'|jj'|nn')
	= \int(dp)\; e^{+i(i-i')p_x}e^{+i(j-j')p_y}e^{+ i(n-n')p_z}\;
	M_{\mu\nu}(p_x,p_y,p_z).
\end{equation}
 And the momentum space $3\times3$ matrix $M(p_x, p_y, p_z)$ looks like
\begin{equation}
M       = \epsilon^{-2} M_1 + M_2,
\end{equation}
\begin{eqnarray*}
M_1     & = & \left( \begin{array}{ccc}
		\Delta(p_y) & -\nabla(p_x)\nabla^*(p_y) & 0 \\
		-\nabla^*(p_x)\nabla(p_y) & \Delta(p_x) & 0 \\
		0 & 0 & 0
	\end{array} \right) \\
M_2     & = & \left( \begin{array}{ccc}
		\Delta(p_z) & 0 & -\nabla(p_x)\nabla^*(p_z) \\
		0 & 2\Delta(p_z) & -\nabla(p_y)\nabla^*(p_z) \\
		-\nabla^*(p_x)\nabla(p_z) & -\nabla^*(p_y)\nabla(p_z) & \Delta(p_x,p_y)
	\end{array} \right).
\end{eqnarray*}

We need eigenvalues and eigenvectors of $M(p)$. Below the
eigensystem follows:
\begin{equation}
\omega_0        = 0,\;
\omega_1        = \Delta(p_x,p_y,p_z),\;
\omega_2        = \epsilon^{-2} \Delta_\epsilon(p_x,p_y,p_z) ,
\end{equation}
\begin{equation}
\left|\:v_0\:\right\rangle      \propto \left( \begin{array}{c}
				\nabla(p_x) \\ \nabla(p_y) \\ \nabla(p_z)
			\end{array} \right),\;
\left|\:v_1\:\right\rangle      \propto \left( \begin{array}{c}
				\nabla(p_x)\nabla^*(p_z) \\
				\nabla(p_y)\nabla^*(p_z) \\
				-\Delta(p_x,p_y)
			\end{array} \right),\;
\left|\:v_2\:\right\rangle      \propto \left( \begin{array}{c}
				+\nabla^*(p_y) \\
				-\nabla^*(p_x) \\
				0
			\end{array} \right)
\end{equation}
where $\left|\:v_0\:\right\rangle$, $\left|\:v_1\:\right\rangle$ and
$\left|\:v_2\:\right\rangle$ are to be
normalized.

\section{}\label{sec:asymptotic}
In this appendix, we prove the asymptotic behavior in Eqn.(\ref{eqn:asymU}) of
$U(2r)$ for the limit $r \gg 1$.
At first, we show that the function $F_\epsilon(\alpha)$ in
Eqn.(\ref{eqn:funcF})
vanishes with $\alpha\to0$ as
\begin{equation}
F_\epsilon(\alpha) \leq A \alpha^2|\ln\alpha|,\
( A > 0 ).
\end{equation}
And then, we show that for $U(2r)$, the correction to the linear
behavior in $r$ is at most ${\cal O}(1/r)$.

Resorting to the mean value theorem in the interval
$[0, \alpha]$, there exists $0 < c < \alpha$ such that
\[
F_\epsilon(\alpha) = F_\epsilon(0) + \alpha\frac{d}{dc}F_\epsilon(c).
\]
Then the inequality follows:
\begin{eqnarray*}
\lefteqn{
F_{\epsilon}(\alpha) = \alpha c \int_0^1 dt\;\left[
	\frac{\sqrt{1-t^2}}{(c^2 + 1-t^2)^{3/2}} +
	\frac{\sqrt{1-t^2}}{(c^2 + \epsilon^2 + 1-t^2)^{3/2}}
\right]
}       \\
	& \leq & \alpha c \int_0^1 dt\; \left(
		\frac{1}{c^2 + 1-t^2} + \frac{1}{c^2 + \epsilon^2 + 1-t^2}
	\right)  \\
	& \leq & \alpha c \left(
		\frac{1}{2\sqrt{1+c^2}}\ln\frac{\sqrt{1+c^2} + 1}{\sqrt{1+c^2}-1}
	\right),\ (0 < c < \alpha \ll \epsilon )  \\
	& \leq & \alpha^2 \left(
	\frac{1}{2\sqrt{1+\alpha^2}}\ln\frac{\sqrt{1+\alpha^2} +
1}{\sqrt{1+\alpha^2}-1}
    \right) \\
	& \sim & \alpha^2|\ln\alpha|
\end{eqnarray*}

Now we decompose $G(\alpha)\equiv F_\epsilon(\alpha) / \alpha^2$ into diverging
part $G_{div}(\alpha)$
and smooth part $G_{smth}(\alpha)$.
Then the above argument shows that the divergence of $G_{div}(\alpha)$ as
$\alpha\to0$ is at most
logarithmic. In consequence, the integration
\[
\int_0^{\pi/2} d\varphi\; \sin^2(2ry)\;\left| \ln\sin \varphi \right|
	= \frac{\pi}{4}\ln 2 - \frac{\pi}{8}\cdot\frac{1}{2r}
\]
leads to the simple asymptotic form
\begin{equation}
\int_0^{\pi/2}d\varphi\;\sin^2(2r\varphi) G_{div}(\sin \varphi) = \mbox{Const.}
+ {\cal O}(1/r),\
	(r \gg 1).
\end{equation}
Finally noting that
\begin{equation}
\int_0^{\pi/2}d\varphi\; \sin^2(2r\varphi) G_{smth}(\sin \varphi)
	= \mbox{Const.} + {\cal O}(1/r),\ (r \gg 1)
\end{equation}
completes the proof.

\begin{figure} 

\vspace{1cm}
\caption{
	Vortex-anti vortex pair. In a three dimensional lattice,
	a vortex line is either infinitely long or a closed loop.
	For highly anisotropic 3D XY-model, however, the above configurations
	is most the important.
	This configuration including the Josephson string is called
	a vortex-antivortex pair.
	(Minnhagen and Olsson, 1991)
}
\label{fig:vortexloop}
\end{figure}

\begin{figure} 

\vspace{1cm}
\caption{
	Dual lattice of cubic lattice. The auxiliary field $\ell$
	is defined at each of the face center in the cubic lattice.
	The constraint of Eqn.(6) to the field $m$
	is automatically satisfied for any value of the field $\ell$.
}
\label{fig:duallattice}
\end{figure}

\begin{figure} 

\vspace{1cm}
\caption{
	Numerical evaluation of the interaction energy. The part (a)
	shows linear dependence on $2r$ for $r \gg 1$.
	At the part (b), the crossover from logarithmic behavior
	for small $r$ to linear one for large $r$ is clear.
	For a comparison, the decoupled case $(\epsilon = 0)$ is
	also plotted with filled circles.
}
\label{fig:intplot}
\end{figure}


\begin{references}
\bibitem {XY:footnote} Hikami and Tsuneto assumed that
     $U(r) \simeq K\ln r + \epsilon^2 K r^2 + 2\mu$;
     quadratic instead of linear for large $r$.
     We note, however, that the
     cutoff factor is the same.
\end{references}
\end{document}